\documentclass{article}

\usepackage[english]{babel}
\usepackage[utf8]{inputenc}
\usepackage{amsthm}
\usepackage{amsmath}
\usepackage{float}
\usepackage{natbib}
\usepackage[margin=1.0in]{geometry}
\usepackage[colorlinks,citecolor=blue,urlcolor=blue,filecolor=blue]{hyperref}
\usepackage{graphicx}
\usepackage{setspace}
\usepackage{amsfonts}
\usepackage[ruled,vlined]{algorithm2e}
\usepackage[title]{appendix} 

\SetKwInput{KwInput}{Input}
\SetKwInput{KwOutput}{Output}
\graphicspath{ {./images/} }

\begin{document}

\title{
A News-based Machine Learning Model for Adaptive Asset Pricing
\thanks{We thank the editors, referees, and all the support from Cornell University.}
}
\author{
Liao Zhu\thanks{Department of Statistics and Data Science, Cornell University, Ithaca, New York 14853, USA.} ,
Haoxuan Wu\thanks{Department of Statistics and Data Science, Cornell University, Ithaca, New York 14853, USA.} ,
Martin T. Wells\thanks{Charles A. Alexander Professor of Statistical Sciences, Department
of Statistics and Data Science, Cornell University, Ithaca, New York 14853, USA.
}}

\date{}
\maketitle

\begin{abstract}
The paper proposes a new asset pricing model -- the News Embedding UMAP Selection (NEUS) model, to explain and predict the stock returns based on the financial news. Using a combination of various machine learning algorithms, we first derive a company embedding vector for each basis asset from the financial news. Then we obtain a collection of the basis assets based on their company embedding. After that for each stock, we select the basis assets to explain and predict the stock return with high-dimensional statistical methods. The new model is shown to have a significantly better fitting and prediction power than the Fama-French 5-factor model.
\end{abstract}


\section{Introduction}
The paper proposes a new asset pricing model -- the \textbf{\emph{News Embedding UMAP Selection (NEUS) model}}, to explain and predict the stock returns based on the financial news. The proposed model is built on top of the recent achievements in asset pricing and natural language processing.

From the asset pricing perspective, the NEUS model is based on the Adaptive Multi-Factor (AMF) model proposed by \cite{zhu2020high}, which provides a modern and more general framework for multi-factor models. The AMF model contains the traditional well-known Fama-French 5-factor model (FF5) \cite{fama2015five} as a special case. The finance theory behind the AMF model is the Generalized Arbitrage Pricing Theory (GAPT) proposed in \cite{jarrow2016positive} and \cite{jarrow2016bubbles} as a modern and more general framework of the traditional Arbitrage Pricing Theory (APT) proposed by \cite{ross1976arbitrage}. The AMF model relaxes the assumption of traditional asset pricing models (such as FF5 model) that the number of factors is small. More importantly, the AMF model uses ``basis assets'' as a ``basis representation'' of all the \emph{realized} risk factors, which includes the traditional \emph{expected} risk factors as special cases. Therefore, the AMF also makes it possible for the adaptive basis asset selection which the traditional multi-factor models do not. Details can be found in \cite{zhu2020high}.

From the aspect of natural language processing and machine learning, the NEUS model is based on the recent progresses including Uniform Manifold Approximation and Projection (UMAP) (\cite{mcinnes2018umap}), Word Embedding, Paragraph Model (\cite{le2014distributed}), Minimax Concave Penalty (MCP) (\cite{zhang2010nearly}), Minimax Prototype Clustering (MPC) (\cite{bien2011hierarchical}), etc.

With the advantages from the two perspectives mentioned above, the NEUS model is shown to have significantly better fitting and prediction power than benchmark FF5 model. We first obtain a collection of the basis assets based on their \textbf{\emph{company embeddings}} derived from the financial news using UMAP, paragraph model and word-embedding. Then for each stock, we select the basis assets to explain and predict the stock return with high-dimensional statistical methods including MPC and MCP, etc. An overview of the NEUS model can be found in Algorithm~\ref{algo_NEUS}. The detail of the NEUS algorithm is in Section~\ref{sec_methodology}.

An outline of this paper is as follows. Section~\ref{sec_methodology} gives the details of the NEUS model. Section~\ref{sec_results} gives the results, including the intercept test results, explanatory results, predictive results, and industry-wise comparison. The NEUS model outperforms the FF5 model in both explanation and prediction, and the out-performance is robust in all industries. Finally, Section~\ref{sec_conclusion} concludes.

\section{Methodology}
\label{sec_methodology}

The NEUS model consists of two key steps. First, we identify a set of basis assets that are representatives of the impact of the market news from a large collection of Exchange Traded Funds (ETFs) based on their company embedding vectors derived from financial news related to them. Then for each stock, we select basis assets that are significantly related to it using a combination of high-dimensional statistical methods. Then we used the basis assets selected to fit an Adaptive Multi-Factor (AMF) model to evaluate the performance of the models. The following subsections will detail each of the steps.

\subsection{Company Vector and Clustering}
For the first part of the analysis, we wish to extract a set of basis assets that form a representative basis for the market. Following similar rationale as seen in \citet{zhu2020high}, we choose the Exchange Traded Funds (ETFs) as the representative source as they aggregate information from a diverse set of products on the market.

We use the financial news text data from the Dow Jones ``Data, News and Analytics (DNA)'' database \url{https://www.dowjones.com/news-you-can-use/}, which contains nearly all articles from the leading financial publishers such as the Wall Street Journal, New York Times, etc. Each article is tagged by the companies and industries that it is related to. We gathered all the articles related to any ETFs and use them in the following analysis.

The data consists of news articles about all relevant ETFs from 2013-2018. We pulled news articles from the Dow Jones news database, selecting from a range of credible sources including the Wallstreet Journal, the Economist and the New York Times. The details of the dataset will be given in the appendix. From the dataset, we selected a set of 603 ETFs that have at least 5 articles in that time-period. The cutoff threshold of 5 is chosen so that there are sufficient news information to represent each ETFs while maintaining a large number of ETFs for basis selection. To maintain consistency, for each of the 603 ETFs, we randomly selected 5 articles from the time-period and concatenated the articles together to form as a news representation.

Next, we learn one vector for each ETF using word embedding with the paragraph model approach \citep{le2014distributed} and call this vector the \emph{\textbf{company embedding}} of this ETF. The paragraph model approach learns one vector for each document and embeds the document in the same space as the word embedding. By utilizing this approach, we can derive a word embedding for each ETF, aka, the company embedding, based on the news articles. Then we can identify a set of basis by clustering the company embeddings. For the implementation, we use doc2vec as a part of the Gensim package \citep{rehurek2010software} in Python.

After combining the articles into one news string for each ETF, we convert all words into lowercase and remove punctuation, numbers and stopwords. Stopwords are common words used in the English language such as "the" and "a" that do not contribute much to meaning of the underlying text. Next, we tokenize each word in the news representation and feed the inputs into a doc2vec model. Following the recommendation of \citet{lau2016empirical}, we utilize a distributed bag-of-words representation where the paragraph vector and neighboring words are utilized to predict context words. We tune three hyper-parameters in the set-up including word embedding vector size ($\{50, 100, 300, 500\}$, window size ($\{5, 10, 15, 20\}$ and epochs $\{50, 100, 200, 400\}$. The optimal hyper-parameters found were vector size of 100, window size of 5 and epoch of 50. We also utilize a minimum count of 3, sub-sampling rate of $10^{-5}$ and negative sample of 5 as recommended by \citet{lau2016empirical}. This set-up provides the most consistent clustering of ETFs.

The next step of the analysis involves clustering the ETFs based on the document vectors and extracting a set of prototypes from the clusters. For this step, we first perform a project of the word embedding from dimension 100 to dimension 5 using Uniform Manifold Approximation and Projection (UMAP) \citep{mcinnes2018umap}. UMAP utilizes cosine distance between nearest neighbors to perform project and has shown to preserve the global and local structure within high-dimensional data. Recent work \citep{zhu2020high} has shown UMAP to be useful for topic extraction in context of document vectors. After projecting to 5 dimensions, we utilize minimax prototype clustering \citep{bien2011hierarchical} in order to derive the clusters and corresponding prototypes. Minimax prototype clustering ensures that the center for each cluster is one of the ETFs, allowing for each basis selection. We experiment with different cutoff threshold for the height of dendrogram and found that a low cut-off threshold of 0.25 performs the best. For the next step, we extracted a set of 147 ETF prototypes as a set of basis for prediction and explanation. The full list of selected ETFs are shown in the appendix.

\subsection{Basis Assets Subset Selection}
After extracting a basis of 147 ETF prototypes from the news dataset, we utilize the Adaptive Multi-Factor model in order to explain and predict the movements of assets in the market. We utilize weekly return data of 3935 trading assets in the marketplace from January 2013 to December 2019. The time-period is subdivided into 5 years of training from 2013 to 2017, one year of validation for year 2018 and one year of testing for year 2019. By dividing the data this way, we prevent look-ahead in the training data and ensures the model tests on out-of-sample data with previously unseen information. For the experiment, we analyze both the explanatory power and predictive power of the news basis and compare our method against the baseline of 5-factor Fama French model. Furthermore, we perform zero-alpha test to test the principle of no-arbitrage.

For each trading asset, denote its return to be $R_i = \{R_{1i}, ..., R_{Ti}\}$ for weeks $t = 1, ..., T$, we will extract a subset of the prototypes that has strong correlation with the return of the asset. We will denote the market return by $P_0$ and return for each of the prototypes by $P_1, ..., P_{p}$ for $p = 147$. Using the Multi-Factor Asset Model, the return can be written as:
$$
(R_{it} - r_f) = \sum_{j=1}^p \beta_{ij} (P_{jt} - r_f) + \epsilon_{it}, \qquad
\epsilon_{it} \sim N(0, \sigma_i^2).
$$
where $r_f$ is the risk-free rate and the coefficients $\beta_i$ and error variance $\sigma_i^2$ are assumed to be constant across time.

In order to select the subset of prototypes that affect each trading asset, we use a sparsity inducing Minimax Concave Penalty (MCP). We chose MCP as it has been shown to have better performance in comparison to the LASSO \citep{zhang2010nearly}; we compare the results with the LASSO in section 4. The selection process is done on the training portion of thee data, with a focus on selecting a set of prototypes that correspond to return of the asset independent from the market. As market return is the primary driver behind movement of returns in trading assets, we first regress the asset as well as all prototypes against the market to get the residuals. We will use $\Tilde{R}_i$ to denote residuals for asset $i$; the residual can be calculated as follows:
$$
(R_{it} - r_f) = \alpha_{0i} (P_{0t} - r_f) + \Tilde{R}_{it}
$$
where $\alpha_i$ is the least squared solution to the regression problem. We perform the same regression to get the residuals for each of the prototypes, denoted by $\Tilde{P}_{i}$.

In order to select the subset of $\{j | \beta_{ij} \neq 0\}$, we minimize the following objective function:
$$
L_\lambda(x_1, ..., x_p) = \min_ ||\Tilde{R}_{i} - \sum_{i=1}^p x_i \Tilde{P}_i||_2^2 + \sum_{i=1}^p q_a(x_i; \lambda)
$$
where a is a hyper-parameter (set to be 3 in simulations) and the penalty function can be written as:
\[
q_a(x; \lambda) = \begin{cases}
\lambda |x| - \frac{x^2}{2a} & \text{if } |x| \le a \lambda \\
\frac{1}{2}a \lambda^2 & \text{if } |x| > a \lambda
\end{cases}
\]

In comparison to the LASSO, this penalty has a flatter tail and penalizes less for large values of $x_i$. This objective function returns a subset of coefficients for each value of $\lambda$. In order to select the value for $\lambda$, we utilize forward validation. Following the work of \citep{friedman2010regularization}, we first utilize MCP to identify a path of $\lambda$ that starts with $\lambda_{\max}$ which is the smallest value that sets all coefficient to 0, then decreases based on log-increments. Due to the sequential nature of time series analysis, we divide the weekly return data into 5 years for training (2013-2017), one year of validation (2018) and one year for testing (2019). For validation, we utilize a rolling window approach previously seen in \citet{nicholson2014high}, \citet{banbura2010large} and \citet{song2011large}. For each week $t$ in the validation, we retrain the model using data from the previous 5 years $(t-1, ..., t-260)$ to predict result for week $t$. The mean-squared error is computed for each the $\lambda$ along the path. We then use the ``1se rule'' to select the $\lambda$. The ``1se rule'' selects the maximum $\lambda$ that has MSE within one standard deviation of the lowest MSE along the validation path, in order to avoid over-fitting. To further prevent over-fitting, we also cap the number of coefficients to be selected to be 20.

\vspace{0.3in}
\begin{algorithm}[H]
\SetAlgoLined
\KwInput{Set of news articles $(N_1, ..., N_m)$ and stock returns $(Y_1, ..., Y_r)$} \
\pmb{Step 1: Paragraph Vector Encoding/Clustering} \
\begin{enumerate}
    \item Aggregate all relevant articles for each ETF of interest
    \item Convert to lower case, remove punctuation/stopwords and tokenize for each ETF
    \item Utilize paragraph vector model to learn one vector for each paragraph
    \item Project paragraph embedding to lower dimension using UMAP
    \item Use minimax prototype clustering to extract key clusters and prototypes as the set of basis for prediction/explanation
\end{enumerate}
\pmb{Step 2: Subset Selection} (for each stock $Y_i = (y_{i, 1}, ..., y_{i, T})$)
\begin{enumerate}
    \item Separate the time-period into training $1,..., t_l$, validation $t_{l}+1, ..., t_v$ and testing $t_{v}+1, ..., T$
    \item Regress out market return from the stock and each the ETFs in the training period
    \item Use MCP on the residuals from step 2 to identify a path of variable selection of basis
    \item Test different basis sets on the validation period using forward validation to calculate MSE for different set of relevant basis
    \item Use 1se rule to select the a sparse set of relevant basis for prediction
    \item Fit linear regression using relevant basis for testing period
\end{enumerate}
\caption{NEUS Algorithm}
\label{algo_NEUS}
\end{algorithm}
\vspace{0.3in}

In this way, we finally derive the estimation (or prediction) of the returns of each stock from the NEUS model. We then compare the goodness of fit between the NEUS model and FF5 model in the Section~\ref{sec_results}. The summary of the NEUS algorithm is in the Algorithm~\ref{algo_NEUS}. Some of the related works can be seen in \cite{jarrow2021low}, \cite{zhu2020adaptive}, \cite{huang2019identifying, huang2020time, huang2021staying} \cite{hu2021revealing}, \cite{lu2017detrending}, \cite{zhang2021form, du2020multiple}, \cite{xu2019u, xu2020small, xu2020improved}, \cite{zhao2020online}, \cite{kong2018study}, \cite{zhu2020high, zhu2021time}.

\section{Results}
\label{sec_results}

From the paragraph vector encoding/clustering, we selected a set of 147 ETFs that are representative of the basis assets in the market. For the analysis, we collected weekly return data of 3692 stocks in the market place. NEUS is then ran on each of the stocks to identify the set of associated basis assets. After selecting the set of basis assets, we then ran simple linear regression to evaluate its out-of-sample explanatory and predictive power in the testing period.

\subsection{Intercept Test}
In order to assess the validity of the adaptive multi-factor model, we will test on the no-arbitrage (no-intercept) assumption. In other words, we will use the intercept to test the existence of arbitrage in our model. First, we will rewrite the model as follows:
$$
(R_{it} - r_f) = \alpha + \sum_{j=1}^p \beta_{ij} (P_{jt} - r_f) + \epsilon_{it}, \qquad
\epsilon_{it} \sim N(0, \sigma_i^2).
$$
where $\alpha$ is the intercept. Next, we will run linear-regression over the 5 year training period for each of the 3692 stocks and extract the p-value for the intercept. A significant intercept indicates the existence of an arbitrage opportunity. We perform a similar analysis for the Fama-French 5-factor (FF5) model as a baseline to compare the results.

\begin{table}
\caption{Intercept Test Results}
\label{tabit}
\centering
\begin{tabular}{ c c c c c}
\hline
P-Value & NEUS & FF5 & NEUS FDR q-val & FF5 FDR q-val\\
 \hline
0-0.05 & $5.74\%$ & $7.34\%$ & $0\%$ & $0.0005\%$\\
\hline
0.05-0.9 & $83.77\%$ & $83.26\%$ & $0\%$  & $0\%$  \\
\hline
0.9-1 & $10.49\%$ & $9.40\%$ & $100\%$ & $99.9995\%$ \\
\hline
\end{tabular}
\begin{flushleft}
{\small Comparison of performances between NEUS and FF5 models in terms of significant intercepts. The results after controlling for false discovery rate using BHY procedure is shown in the last two columns. }
\end{flushleft}
\end{table}

As seen in Table \ref{tabit}, for the NEUS model, 5.74\% of the stocks had an significant intercept p-value at the $0.05$ threshold. For the FF5 model, 7.34\% of the stocks had an significant intercept p-value at the $0.05$ threshold. NEUS slightly outperforms the FF5 since fewer stocks has a significant intercept. After correcting for false discovery rate using Benjamini-Hochberg-Yekutieli \cite{benjamini2001control} correction similar to \cite{zhu2020high}, we found that both NEUS and FDR have a FDR rate close to 0. Therefore, both models are consistent with the no-arbitrage assumption.

\subsection{Explanatory Analysis}
\begin{figure}[h!]
  \centering
  \caption{Adjusted R-Squared for Explanation Comparison between FF5 vs GNU Models}
  \label{figINEX}
  \vspace{5pt}
  \begin{tabular}{c c }
  In-sample & Out-of-sample \\
  \includegraphics[scale=.5]{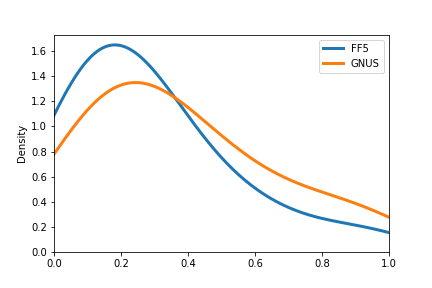} & \includegraphics[scale=.5]{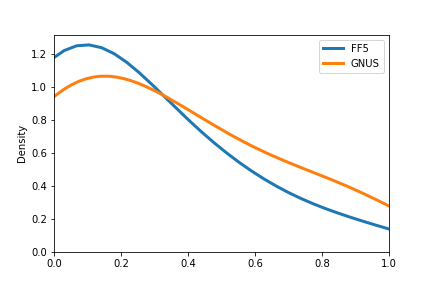}
  \end{tabular}
  \begin{flushleft}
   {\footnotesize The plot illustrate the density adjusted R-squared for FF5 vs NEUS for in-sample explanation and out-of-sample explanation comparisons. As seen, NEUS outperforms FF5 in both in-sample and out-of-sample explanation. }
   \end{flushleft}
\end{figure}

The first analysis involves understanding the usefulness for the set of basis vectors in explaining returns of the stocks in the marketplace. For this analysis, we first compare the in-sample adjusted R-squared between 5 factor Fama French model and the NEUS algorithm. We chose adjusted R-squared as the main metric for comparison as it corrects for the number of explanatory variables. Table \ref{tab1} details the results. The mean adjusted R-squared for the NEUS model is 0.375 while the mean adjusted R-squared for the FF5 model is 0.287. Figure \ref{figINEX} shows plot of density of results for the two models. Overall, the NEUS algorithm shown an improvement of $30.7\%$ in terms of in-sample performance. For the selection of basis, 61 out of 147 basis are selected at least once for explanation.

Next, following the analysis seen in \citep{zhu2020high}, we perform a F-test in order to see if there is a significant difference between the goodness-of-fit for FF5 model and the NEUS model for each individual stock. We use a F-test to compare the SSE between the FF5 model and the FF5 + NEUS model where FF5 + NEUS model has predictors from both the Fama French and the ETFs selected by the NEUS algorithm. With this set-up, we can see that the FF5 model has a degree of freedom equal to $n - r_1$ where $r_1 = 5$ and the FF5 + NEUS model has degree of freedom equal to $n - r_1 - r_2$ where $r_2$ is the number of ETFs selected by the NEUS algorithm. The F-statistics can be derived as follows:
$$
\frac{SS_{F} - SS_{G} / r_2}{SS_G / (n - r_1 - r_2)} \sim F_{r_2, n - r_1 - r_2}
$$
where $SS_F$ is the sum of residuals for the FF5 model and $SS_G$ is the sum of residuals for the FF5 + NEUS model. Out of the 3692 stocks, 3141 selected at least one of ETFs as a significant predictor for the return. Out of the 3141 stocks, 2948 ($93.85\%$) of them have significant F-score exceeding the $0.05$ threshold. In order to correct to multiple testing error, we will use two types of correction. First, with Bonferroni correction, 2079 ($56.38\%$) of the stocks have F-scores exceeding the threshold $\frac{0.05}{3692}$. Second, since Bonferroni correction is quite a harsh correction, we also use Benjamini-Hochberg procedure to control for False Discovery Rate. Using a FDR threshold of 0.05, we find that 2947 ($93.80\%$) of the stocks have significant F-scores. From this analysis, we can see that adding the ETFs improve the performance in explanation of returns for a majority of stocks.

\begin{figure}[h!]
  \centering
  \caption{Number of Predictors Selected for Explanation}
  \label{figNoVar}
  \includegraphics[scale=.7]{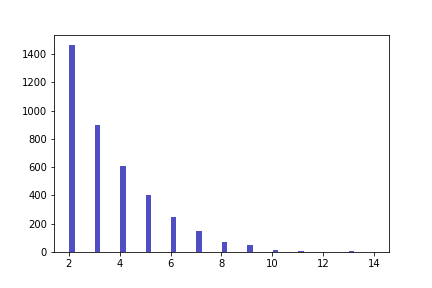}
  \begin{flushleft}
   {\footnotesize The plot illustrates histogram of count of stocks with varying number of predictors selected for explaining returns. As seen, the majority of stocks require less than 5 predictors for explanation. }
   \end{flushleft}
\end{figure}

For out-of-sample explanation, we compared the out-of-sample adjusted R-squared value for both NEUS model and the FF5 model on the one year testing period. A rolling window strategy is used once again to predict out-of-sample one week at a time. The out-of-sample mean adjusted R-squared for the FF5 is 0.199 with a standard error of 0.005. The out-of-sample mean adjusted R-squared for the NEUS is 0.308 with a standard error of 0.006. Figure \ref{figINEX} illustrates the density plot for the two models. The NEUS model outperforms FF5 model by 54.5$\%$, supporting the idea that the 5 Factor Fama French model overfits the data. The NEUS model shows a much smaller decrease from in-sample adjusted R-squared to out-of-sample adjusted R-squared, showing that robustness of the MCP penalty and rolling window cross-validation approach in selecting the appropriate number of predictors.

\subsection{Predictive Analysis}
\begin{table*}
\caption{Comparison of NEUS vs FF5}
\label{tab1}
\centering
\begin{tabular}{c c c}
\hline
 & NEUS & FF5\\
 \hline
In-sample Explanation & $0.375_{(0.004)}$ & $0.287_{(0.004)}$ \\
\hline
Out-of-sample Explanation & $0.308_{(0.006)}$ & $0.199_{(0.005)}$ \\
\hline
Out-of-sample Prediction & $0.018_{(0.0002)}$ & $-0.027_{(0.002)}$ \\
\hline
\end{tabular}
\begin{flushleft}
{\small Comparison of performances between NEUS and FF5 models for explanation and predictions of stocks in the market place. Mean adjusted R-squared are shown in the tables with standard error in subscript.}
\end{flushleft}
\end{table*}
The second analysis involves understanding the usefulness of the ETF basis for predicting future returns of stocks. For this analysis, we use one week ahead prediction using returns of the ETFs from the previous weeks. A similar training/rolling window validation method is used to select the set of relevant basis for NEUS algorithm. And once again, we compare the results against the FF5 model as baseline. We utilize out-of-sample adjusted R-squared as a baseline for comparison. For calculation of out-of-sample adjusted R-squared, we utilize the mean of the validation period as the baseline for the denominator. In terms of adjusted R-squared, the NEUS algorithm achieved an average out-of-sample adjusted R-squared of 0.18. In comparison, the Fama French model baseline achieves an out-of-sample adjusted R-squared of around -0.029. The result for the 5 factor Fama French model is to be expected as due to the prevalence of the model, it no longer has any predictive power in the marketplace. As seen, the NEUS algorithm is able to select a good set of basis that have a good predictive power in the market.

\begin{figure}[ht]
  \centering
  \includegraphics[scale=.35]{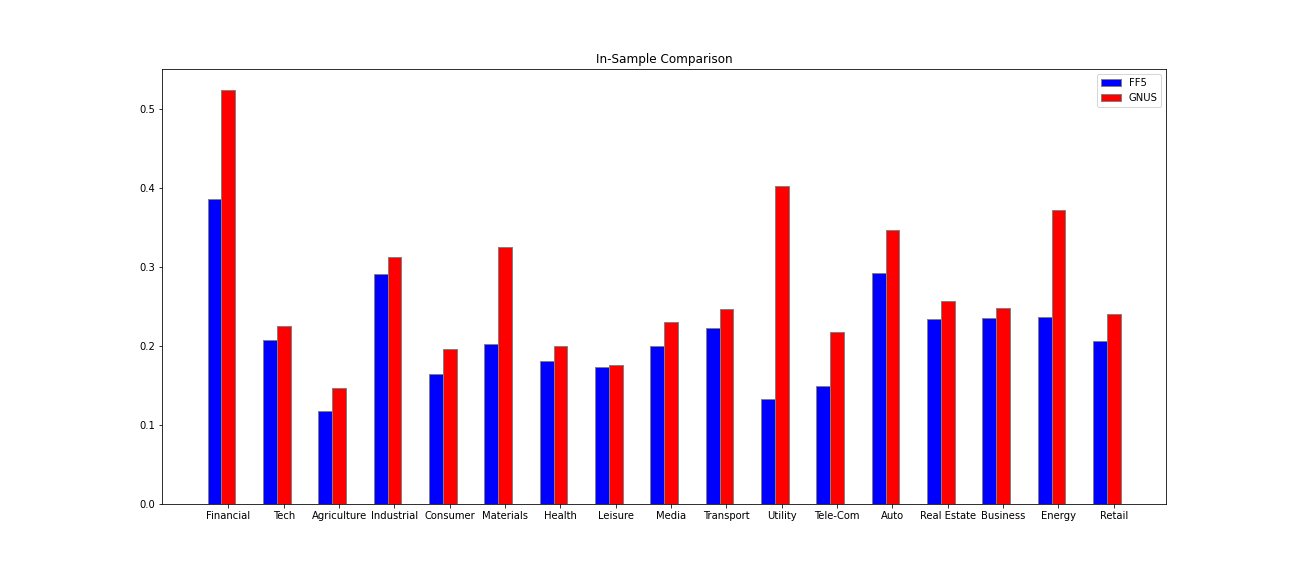}
  \includegraphics[scale=.35]{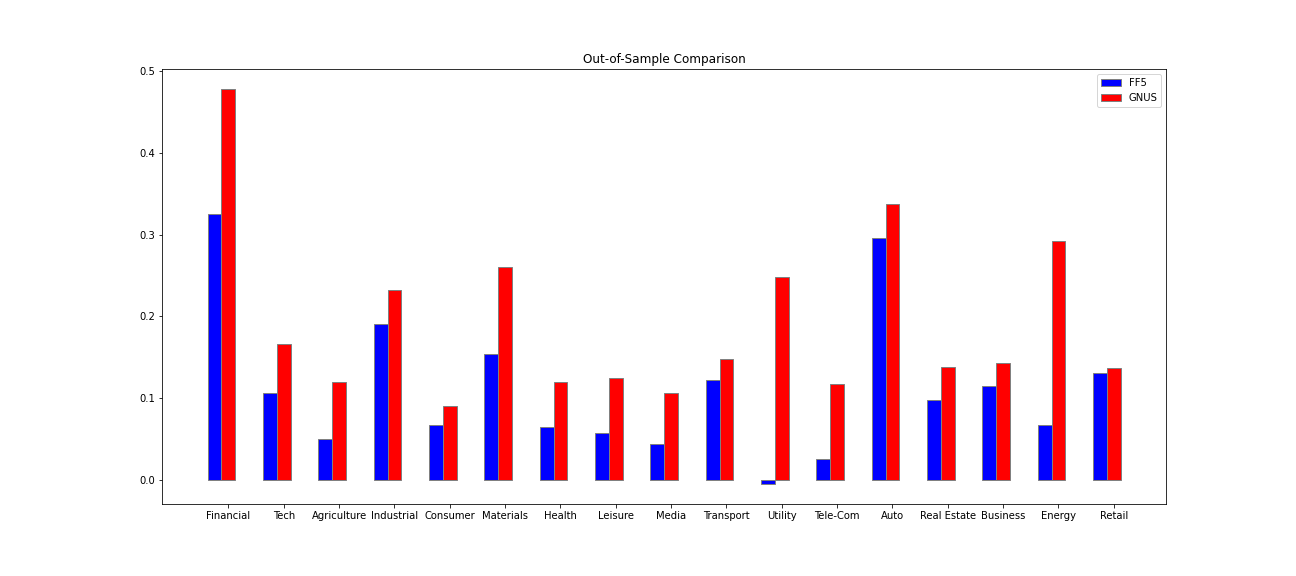}
  \caption{\label{figInd} Comparison of average adjusted R-squared for in-sample and out-of-sample explanation for different industries between FF5 model and NEUS model. }
\end{figure}

As seen in the figure, for the in-sample performances, the NEUS  model slightly out-performs the FF5 model in almost all industries. There is not one industry in which the FF5 model outperforms the NEUS model, showing the robustness of the NEUS model and its ability to provide better basis for all sectors of the marketplace. The NEUS model significantly outperforms the FF5 baseline in the financial sector, utility sector and energy sector. A similar pattern can be seen in the out-of-sample comparison. Once again, the FF5 model does not outperform the NEUS model in any of the industries. The NEUS provide a higher baseline level of mean-adjusted R-squared for all industries; in comparison, the FF5 model heavily struggles to explain movement in the Utilities and Telecommunication Services industries.

\subsection{Industry-wise Comparison}
To better understand the performance of NEUS model in terms of explaining the returns of assets in the marketplace, we looked more in-depth at the mean adjusted R-squared for each industry. We separated the assets into industries based on classification from the Dow Jones databases. At the highest level, the market place is divided into 17 industries: 'Financial Services', 'Technology', 'Agriculture', 'Industrial Goods', 'Consumer Goods', 'Basic Materials/Resources', 'Health Care/Life Sciences', 'Leisure/Arts/Hospitality', 'Media/Entertainment', 'Transportation/Logistics', 'Utilities', 'Telecommunication Services', 'Automotive', 'Real Estate/Construction',
'Business/Consumer Services', 'Energy', and 'Retail/Wholesale'. Each asset is classified into one of the following as their primary industry. We compared the results of the FF5 model vs the NEUS model on mean adjusted R-squared in-sample and out-of-sample for each of the industries. The result can be seen in figure \ref{figInd}.

\section{Conclusion}
\label{sec_conclusion}

The paper proposes a new asset pricing model -- the News Embedding UMAP Selection (NEUS) model, to explain and predict the stock returns based on the financial news, using a combinations of various machine learning models (word-embedding, paragraph model, UMAP, MPC, MCP, etc.). For each basis asset, we first derive its company embedding vector based on its financial news. Then we obtain a collection of the basis assets based on their company embeddings. After that for each stock, we select the basis assets to explain and predict the stock return with high-dimensional statistical methods. The new model is shown to have a significantly better fitting and prediction power than the Fama-French 5-factor model (FF5) in all aspects.

As shown in the results, the mean adjusted R-squared for GNUS (0.375) is 31\% higher than the FF5 (0.287) for explanation. But more importantly, the out-of-sample mean adjusted R-squared for GNUS (0.308) is 55\% higher than the FF5 (0.199) for simultaneous-time explanation. This clearly shows that the GNUS out-performs the FF5 in explanation in both in-sample and out-of-sample. Furthermore, for the 1-week ahead prediction, the out-of-sample mean adjusted R-squared for the GNUS model outperforms the FF5 by 167\%. This provides clear evidence that FF5 is overfitting and that the GNUS model is much more powerful in capturing the potential realized risk factors than the FF5 model.

Finally we do another robustness test -- comparing the performance between GNUS and FF5 in all industries. Again, the GNUS model outperforms the FF5 in all industries in both in-sample and out-of-sample period. There is not even one industry in which the FF5 model outperforms the GNUS model, showing the robustness of the GNUS model and its ability to provide better basis asset for all sectors of the marketplace. The FF5 model heavily struggles to explain any movement in the Utilities and Telecommunication Services industries, and the out-performance of GNUS over FF5 is more extreme in the financial sector, utility sector and energy sector.

\bibliographystyle{plainnat}   
\bibliography{paper}


\end{document}